# Tycho Brahe, Abū Maʿshar, and the *comet beyond Venus* (ninth century A.D.)

Ralph Neuhäuser (AIU, University Jena, Schillergäßchen 2, 07745 Jena, Germany, rne@astro.uni-jena.de),
Paul Kunitzsch (LMU München, Davidstraße 17, 81927 München, Germany),
Markus Mugrauer, Daniela Luge (AIU, University Jena, Schillergäßchen 2, 07745 Jena, Germany),
Rob van Gent (Mathematical Institute, University Utrecht, P.O. Box 80010, 3508 TA Utrecht, The Netherlands)

Abstract.
From his observations of the A.D. 1572 supernova and the A.D. 1577 comet, Tycho Brahe concluded that such transient celestial objects are outside the Earth's atmosphere, and he quoted the 9th century A.D. Persian astrologer and astronomer Abū Maʿshar: *Dixit Albumasar, Cometa supra Venerem visus fuit*, i.e. that he had reported much earlier that comets were seen beyond Venus. However, even from a more detailed Latin translation, the observations and logic behind Abū Maʿshar's conclusion were not understandable. We present here the original Arabic text (MS Ankara, Saib 199) together with our translation and interpretation: Abū Maʿshar reported that he had observed Venus in (or projected onto) the tail of a comet and concluded that the comet was behind Venus, because he had observed the extinction of Venus due to the cometary tail to be negligible (*light of Venus was unimpaired*). He then concluded that the comet would be located behind Venus. He also mentioned that others had observed Jupiter and Saturn in cometary tails, so that those comets would even be located beyond those two outer planets – *in the sphere of the stars*. The dates of the observed close conjunctions were not mentioned; using known orbital elements for a few comets, we found a few close conjunctions between comets and planets from A.D. 770 to 868, but we cannot be sure regarding which conjunctions were reported. While the argument of Abū Maʿshar is not correct (as cometary tails are optically thin), parts of the conclusion – namely that comets are outside the Earth atmosphere and beyond the moon – is correct. This may have helped Tycho Brahe to come to his revolutionary conclusion.

1 Introduction

According to Aristotle's *Meteorology* (book I, chapter 6-7, 342 b 25 to 345 a 7), comets were long thought to be atmospheric phenomena. Tycho Brahe's work about the comet of A.D. 1577 is usually considered to be the first to show convincingly that comets are located outside the Earth atmosphere, even beyond the Moon and among the planets[1]. Tycho Brahe, commenting on the classification of the (super-)nova of 1572 as comet by Adam Ursinus, wrote that he does not want to discuss the astrological meaning of the nova, but quoted Abū Maʿshar indirectly through Adam Ursinus (A.D. 1524-1590) and Jerome Cardanus (A.D. 1501-1576):

"Id solummodo, quod in fine sui Scripti ex Aphorismo quodam Cardani (qui sic habet) citat: *Dixit Albumasar, Cometa supra Venerem visus fuit; non igitur in Elementari Regione est, contra Philosophum.*" [2]

We translate this into English as follows: "Just this, that he [Adam Ursinus], at the end of his work, quotes from a certain sentence of Cardanus (which is written there as follows): Albumasar said: A comet was seen above Venus; it is therefore not in the sphere of the [four] elements, contrary to the philosopher."

The citation was taken literally from Adam Ursinus, who took it from Jerome Cardanus (who has *cometes* instead of *cometa*), there is no additional information available in those two works.

The quoted scholar called by his Latin name *Albumasar* is the Persian astronomer and astrologer Abū Maʿshar, his full name is *Abū Maʿshar Jaʿfar b. Muḥammad b. ʿUmar al-Balkhī*.[3] He was born on A.D. 787 Aug 10 (171 Hijra) in Balkh, Khurāsān[4] (now in Iran), he died on A.D. 886 Mar 9 (272 Hijra)[5]. Abū Maʿshar lived mostly in Baghdad (Iraq) and started studying astronomy when he was already 47 years old – motivated by his colleague al-Kindī.[6] Abū Maʿshar is mainly known for his astrological works, e.g. the *Introduction* (to astrology), translated to Latin by Johannes Hispalensis[7] (John of Seville) and then quoted very often, but he has also written about astronomical observations, e.g. in a Zīj, a list of stars and their parameters[8]. Aristotle's *Meteorology* was translated into Arabic in the second half of the eighth century A.D.[9], so that it was probably known to Abū Maʿshar.

Given the short quotation from Tycho Brahe, it is necessary to find the original Arabic text, on which the quotation (through Cardanus) is based, in order to understand how Abū Maʿshar, seven centuries before Tycho Brahe, concluded that comets are outside the Earth atmosphere, even beyond Venus. While a Latin translation of the relevant text by Abū Maʿshar is available, the logic behind his argument was not clear.[10]

We will present first the problem, namely that the original Arabic text upon which the Latin translation and the quotation by Tycho Brahe are based, was not yet used to understand, how Abū Maʿshar came to his conclusion (Sect. 2). Then, we present the relevant Arabic original text together with our English translation (Sect. 3) and discuss the astronomical interpretation of the text (Sect. 4). In Sect. 5, we calculate and consider which comet – planet conjunctions Abū Maʿshar and his colleagues may have observed, also taking into account historic Chinese and European observations. Finally, in Sect. 6, we conclude with a summary.

2  The problem: Which Arabic work and text was quoted by Tycho Brahe ?

Goldstein[11] already tried to trace the quotation by Tycho Brahe (A.D. 1577) through Cardanus (A.D. 1547) as far back as possible. He quoted a work by Fortunio Licetus (A.D. 1577-1657) [12] as follows:

"The well-known Jerome Cardanus … unanimously affirm that among the Arabs who cultivated astronomy were … Albumazar, who in the year of the Lord 844 observed a comet above the orb of Venus that was certainly larger than Venus. The parallax of the comet was found to be perceptable. Parallax was determined by those astronomers with a most accurate ruler which indicated the altitude of stars." [13]

As found by Thorndike[14], this quotation probably goes back to the following text in a Latin translation of a work by Abū Maʿshar 's student Shādhān often called *Abū Maʿshar in Shādhān*. We cite here from the Latin manuscript Erfurt Amplon. Q.352 fols. 11v-17r (probably 13th/14th century A.D.), which was consulted by us, but not by Thorndike (folio 15r, from line 13 on):

"Dixit Albumasar: Dicunt quidam et ipse Aristoteles quod cometae consistunt in celo in sphera ignium et nihil ex ipsis fit in celo et quia celum non suscipit aliquam passionem. Sed erraverunt omnes circa talem opinionem. Ego enim ipsis oculis vidi cometam super Venerem et sciebam quod cometa erat supra Venerem, quia non immutabat colorem ipsius. Et dixerunt mihi multi, quod ipsi viderunt cometam supra Iovem et alii viderunt supra Saturnum" (with a notice in the right margin saying: "de cometis").

This is fully consistent with Thorndike's translation of other Latin manuscripts:

"Abū Maʿshar said: The philosophers say, and Aristotle himself, that comets are in the sky in the sphere of fire, and that nothing of them is formed in the heavens, and that the heavens undergo no alteration. But they all have erred in this opinion. For I saw with my own eyes a comet beyond Venus, and I knew that the comet was above Venus, because its colour was not affected. And many have told me that they have seen a comet beyond Jupiter and sometimes beyond Saturn." [15]

Federici Vescovini[16] published an edition of the Latin texts of this work, where the text of the relevant paragraph above is fully consistent with the translations by Thorndike[17] and us given above. Thorndike[18] commented this paragraph as a "remarkable antedating of the views of Tycho Brahe" and wrote in the introduction about the work that "it correctly represents comets as celestial phenomena, more distant from the Earth than the planet Venus, or even Jupiter and Saturn."[19]

Apparently, neither Goldstein[20] nor Thorndike[21] consulted any original Arabic manuscript of the work called *Abū Maʿshar in Shādhān* or *Mudhākarāt*[22] (which means "Discussions"), Thorndike[23] consulted two Latin translations from the fourteenth and fifteenth century A.D., namely Bodleian, Oxford, UK, MS Laud. Misc. 594 and BnF, Paris, France, MS 7302. One of the titles of the Arabic work is *Kitāb Abī Maʿshar fī asrār ʿilm al-nujūm*[24] (which means *Book (by) Abū Maʿshar on the secrets of the science of the stars*); it was written by his student Shādhān b. Baḥr and consists mainly of answers by *Abū Maʿshar* to questions by Shādhān and maybe other students on astronomical and astrological topics. Not much is known about that student, *Abū Saʿīd Shādhān b. Baḥr* (for short: Shādhān), an Iranian name, once called *al-Kirmānī*, i.e. a native of the Kirmān province (but not the town Kirmān).[25] Translations of Latin translations are available in parts in English[26] and in full in Italian.[27]

The available Latin translations are shortened and incomplete, and may possibly be based on a Greek or Hebrew translation of the original Arabic[28]. The Greek text survived as Codex Angelicus 29 from the 14th century A.D.[29] and, in parts, as Codex Vaticanus graecus 1056, which both contain the relevant paragraph:[30]

Εἶπον τῷ Ἀπομάσαρ· λέγουσιν οἱ φιλόσοφοι καὶ αὐτὸς ὁ Ἀριστοτέλης ὅτι οἱ κομῆται τοῦ οὐρανοῦ συνίστανται ἐν τῇ τοῦ πυρὸς σφαίρᾳ καὶ οὐδὲ εἷς αὐτῶν ἐν τῷ οὐρανῷ γίνεται καὶ ὅτι ἀνεπίδεκτος ὁ οὐρανός ἐστι τινος πάθους· ἀλλ' ἐσφάλησε πάντη περὶ τὴν τοιαύτην δόξαν, ἐγὼ γὰρ οἰκείοις ὀφθαλμοῖς εἶδον κομήτην ἄνωθεν τῆς

Ἀφροδίτης καὶ ἔγνων ὅτι „εἴδομεν κομήτην ἄνωθεν τοῦ Διὸς καὶ ἔτερον ἄνωθεν τοῦ Κρόνου". (f. 51v)[31]

Our English translation is as follows:

"I said to Albumasar: 'The philosophers argue and Aristotle himself, that the comets of the heaven come together in the sphere of fire and that not anyone of them is formed in the heaven and that the heaven cannot accept any event. But all erred in that regard, because I saw with my own eyes a comet beyond Venus and I learnt that >we saw a comet beyond Jupiter and another one beyond Saturn<'."

As we can see, the text here is said to originate from Shādhān, the student, and not from Abū Ma$^c$shar. He quotes that *"we saw"* comets beyond Jupiter and Saturn. By comparison with the Arabic text, we will see below that this Greek translation is not only highly incomplete, but also wrong in several regards. This is probably due to the fact that the translator did not understand the logical reasoning given by Abū Ma$^c$shar.

From the texts quoted above by both Licetus (*parallax ... perceptable*) and the Latin translation of *Abū Ma$^c$shar in Shādhān* (*its colour was not affected*), it cannot be understood, how Abū Ma$^c$shar concluded that comets are located beyond Venus. Hartner wrote about Thorndike's translation: "... l'interpréter comme le résultat de véritables observations, c'est-à-dire de déterminations des parallaxes variables d'une ou de plusieurs comètes ... mais de la couleur non affectée ou altérée de la comète."[32] Then, Federici Vescovini wrote: "Ancora Abū Ma$^c$shar: ... perché egli stesso ha visto una cometa sopra Venere ed egli sapeva che era sopra Venere perché non mutava il colore di quel pianeta", but again only from a Latin translation.[33] Parallaxes are not known to have been measured by Arabs or others in the ninth century A.D.; and inside the solar system, one cannot conclude on the distance from a colour, because extinction effects are negligible here. While Hartner concluded that the whole discussion by Abū Ma$^c$shar has only an astrological sense, so that the citation by Tycho Brahe would be a misunderstanding[34], Dunlop remarked that "before leaving the question it seemed desirable to recover if possible Abū Ma$^c$shar 's original Arabic".[35]

3  The solution: The original Arabic text in *Abū Ma$^c$shar in Shādhān*

The quotation given above is from the work *Abū Ma$^c$shar in Shādhān.* There are ten Arabic manuscripts of this work listed[36], the oldest one is located in Ankara, Turkey, namely manuscript (MS) Ankara Saib 199 (folios 1a to 26a) dated to the sixth century Hijra[37] (twelfth century A.D.), which should be the complete MS. We have consulted that MS.

MS Saib 199 also includes the work *Kitāb fī dalā'il al-qirānāt wa-l-kusūfāt* by al-Battānī (dated seventh century Hijra[38], on "conjunctions") and a work ascribed to *Kanaka al-Hindī* ("Kanaka, the Indian"), also on conjunctions. All three texts are also available as micro-film and hardcopy in the library of the Institut für Geschichte der Arabisch-Islamischen Wissenschaften, Frankfurt am Main, Germany. Neither the date of the copy nor the name or place of the copyist are mentioned in the manuscript. From the fact that one of the three texts was later dated to the sixth century and one to the seventh century, we conclude that MS Saib 199 is from the sixth or seventh century Hijra (i.e. 12th/13th century A.D.). It is then only slightly younger than the Latin MS Erfurt mentioned above.

We have also consulted MS Paris 6680 (folios 1 to 25) from the twelfth century Hijra,

which is, however, a different work, it was listed as *possibly being Abū Ma<sup>c</sup>shar in Shādhān*[39]. We would also like to note that the relevant parts were previously not found in the Arabic MS Cambridge no. 1028 Gg. 3.19 (folios 1 to 20a, 767 Hijra)[40]. Both MS Esat 1967 in Istanbul's Süleymaniye Library and MS Cairo Ṭal<sup>c</sup>at mīqāt 157 are different works[41]. Pingree[42] consulted three MSS: Huntington 546 in the Bodleian Library, Oxford, the above mentioned MSS Cambridge no. 1028 Gg. 3.19, and Ankara Saib 199, and concluded that only the latter is complete; Pingree compared it to two Greek versions[43] and the Latin edition by Federici Vescovini[44], but neither cited the Arabic text nor gave a translation[45]. We have then also consulted the MS Teheran Millī 1634/10, but it was a different work by Abū Ma<sup>c</sup>shar. We have also tried to consult the manuscript in the Teheran Khunjī library, but were told that it was dislocated to an unspecified location outside Iran. Hence, of the ten MSS listed by Sezgin[46], at least four are different works, one disappeared (Khunjī), and two are incomplete regarding our paragraphs (Oxford and Cambridge).

Sezgin did not only find the Ankara Saib 199 MS, but he also noticed the two relevant paragraphs in the Arabic text by stating:

"Astronomiehistorisch gesehen ist sehr aufschlussreich – wie L. Thorndike und W. Hartner gezeigt haben –, dass sich Tycho Brahe (gestützt auf ein Zitat von Cardano) bei seiner außerordentlich wichtigen Theorie, dass die Kometen sich im Universum frei bewegen ... auf *Abū Ma<sup>c</sup>shar* beruft ... Wie ich es z. Zt. übersehe, scheint der lateinische Text eine ziemlich freie Kombination von zwei Stellen des arabischen Originals zu sein."[47]

Hence, Sezgin already noticed that the Latin text is a free combination of two paragraphs in the Arabic original; he then gives the Arabic text in Latin transcription in his footnote 4 on pages 156 and 157.

We here present the edition of the relevant passages of the Arabic text of *Abū Ma<sup>c</sup>shar in Shādhān* as taken from MS Saib 199 fol. 15b–16a, the original Arabic text is shown here in Fig. 1 (differences to Sezgin's edition are mentioned below, our comments or additions are in square brackets) and reads as follows:

*"Qultu li-Abī Ma<sup>c</sup>shar: al-kayd ma huwa ? Qāla yaqūlu aṣḥāb al-ḥisāb innahu shay' yu'aththiru wa-lā yurā, wa-qad qāla qawm innahu laṭkhatun* (a), *fa-ammā anā fa-lam a<sup>c</sup>rifhā <sup>c</sup>iyānan, wa-qad qāla qawm innahu kawkab munīr janūbī, wa-qāla qawm innahu jawzahr li-ba<sup>c</sup>ḍ al-aflāk; qultu anta ma taqūlu ? Qāla lastu aqna<sup>c</sup>u min dhālika kullihi bi-shay' wa-akrahu ayḍan an adfa<sup>c</sup>a shay'an qad takallama fīhi ahl al-<sup>c</sup>ilm qablī. Wa-qāla wa-yaqūlūna inna sayrahu fī kulli sana darajatān* [MS: *darajatayn*] *wa-niṣf wa-fī kull yawm arba<sup>c</sup> wa-<sup>c</sup>ishrūna* [MS: *arba<sup>c</sup>a wa-<sup>c</sup>ishrīna*] *thāniya wa-fī kull shahr ithnatā <sup>c</sup>ashrata daqīqa".*

(a) Sezgin has *la-najmatun* ("(just) a star").[48]

Then, we continue with MS Ankara Saib 199 fol. 20b - 21a (original in Fig. 2):

*"Qultu li-Abī Ma<sup>c</sup>shar: al-kawkab al-janūbī al-munīr alladhī narāhu fī qismat al-tāsi<sup>c</sup> min al-dalw abyaḍ ka-annahu <sup>c</sup>uṭārid wa-akbar fī ra'y al-<sup>c</sup>ain. Qāla: aẓunnuhu alladhī yuqālu lahu ra's al-jawzā'. Qultu za<sup>c</sup>ama l-mūbadh anna al-Furs tusammīhi al-bārī* [until here not given in Sezgin]; *wa-qāla l-Jayyānī al-munajjim innahu al-kayd. Qāla* [*Abū Ma<sup>c</sup>shar*]: *lā adrī mā aqūlu fī hādhā. Qultu li-Abī Ma<sup>c</sup>shar: al-nayāzik al-ithnā*

[MS: *al-ithnay*] ᶜ*ashar. Qāla: bāṭil, lā ḥaqīqata lahu; al-nayāzik laysat kawākib tusayyaru wa-innamā hiya nayāzik taḥduthu min majran ghayr majārī l-kawākib al-sabᶜa wa-l-aghlab ᶜalā wahmī annahā* (b) *taḥduthu fī falak al-burūj, li-annī ra'aytu al-dhu'āba al-maghribīya allatī ṭalaᶜat mundhu sinīna wa-kānat al-zuhara* (c) *fī ra'y al-ᶜayn maᶜahā fī jasadihā lā ... ᶜanhā* (d)*, wa-lā ... ᶜanhā* (e) *wa-lam* (f) *yaltabis ḍaw'uhā bi-ḍaw'ihi* (g) *wa-lā kānat min* [or: *bayna* ?] *al-nūrayn mumāzaja fa-ᶜalimtu annahā fawqa l-zuhara li-tamām nūr al-zuhara fī ra'y al-ᶜayn; akhbaranī ghayr wāḥid annahum ra'aw zuḥal ka-dhālika maᶜa baᶜḍ al-nayāzik wa-ra'aw al-mushtarī ka-dhālika*".

Sezgin has: (b) *innamā*, (c) *al-zuhra*, (d) *mutabāyina* (?) *īyāhā*, (e) *mutasāyira* (?) *lahā*, (f) *wa-lā*, (g) *bi-ḍaw'ihā*; we are not sure what is written at the location given as "(?)" in Sezgin and by "..." by us.

Our translation of these two paragraphs is as follows (our comments and additions in square brackets):

"I [Shādhān] said to Abū Maᶜshar: What is *al-kayd* ? He [Abū Maᶜshar] said: The scientific astronomers say it is something which produces traces [*yu'aththiru*], but cannot be seen. People have said, it was a [nebulous] spot. But I have not seen it with my own eyes. Some people have argued, it would be a bright southern star; and [other] people have said, it would be the node [*jawzahr*] of certain spheres [like the lunar nodes[49]]. I [Shādhān] said: And what do you [Abū Maᶜshar] say ? He [Abū Maᶜshar] said: I am not happy with any of those; and I also do not like to dismiss what scholars before me have said. And he [Abū Maᶜshar] said: They say, it would move every year by two and a half degrees and every day by 24 seconds and every month by 12 minutes" (Saib 199, folio 15b - 16a).

"I said to Abū Maᶜshar: The bright southern star that we see in the section of the ninth [house?] in Aquarius is white as if it were Mercury or larger when seen with the eye. He [Abū Maᶜshar] said: I think it is the one that is called the "Head of Orion". I [Shādhān] said: The Mobad [Zoroastric priest] said that the Persians call it *al-bārī*. And the astronomer/astrologer al-Jaiyānī said: This is *al-kayd*. He [Abū Maᶜshar] said: I do not know, what I should say on that. I [Shādhān] said to Abū Maᶜshar: [They are] the twelve *nayāzik* [comets]. He [Abū Maᶜshar] said: Nonsense, that is not correct; the *nayāzik* [plural of *nayzak*] are not celestial objects, which orbit in normal spheres, moreover they are celestial objects, which are formed [or: orbit] in an orbit different from the orbits of the seven [planets]. It appears to me most probable that they are formed [or: orbit] in the zodiacal sphere, because I have seen the western [or: evening] comet [*al-dhu'āba* for: lock of hair], which appeared years ago, and for the eye Venus was standing near it in its body [*fī jasadihā*], not distinguishable from it and not moving away from it, and its [Venus'] light did not mix with its [the comet's] light, and there was no blending between the two lights [or: light sources]. Hence, I could conclude that it [the comet] stood above [or: beyond/behind] Venus, because the light of Venus was unimpaired for the eye. More than one told me that they saw Saturn in a similar manner with certain *nayāzik* [comets], and also Jupiter" (Saib 199, folio 20b - 21a).

Some parts of this second paragraph are quite similar to Thorndike's English translation[50] of a Latin translation of *Abū Maᶜshar in Shādhān*, see Sect. 2, even though shortened and slightly altered (e.g. *Aristotle* instead of the *scholars before me*, or *colour* instead of *light*), so that it was hard to understand, while our translation of the original Arabic can readily be understood. The connection between the two paragraphs is given by the fact that both discuss the nature of *al-kayd*.

Let us first comment on some rare words and names in the text:

The word *al-kayd* has an Indian origin, namely from the Sanskrit name *Ketu* for the descending node of the lunar orbit, depicted often as the tail of a dragon, whose head is *Rāhu*, the ascending node[51]; those lunar nodes were early considered as the locations where lunar and solar eclipses can happen. This particular meaning of *al-kayd* was not supported by Abū Maᶜshar. However, the angular velocity given at the end of the first paragraph (2.5 degrees per year) is not the correct angular velocity of the lunar nodes (18.6 yr), but 144 yr. Given the description of *ketu* as the *tail* of a dragon, the object was later considered to be a star with tail (Indian *dhūmaketu* for *smoke-ketu*), i.e. a comet and, then, to be a negative portent given its irregular appearance.[52] This is then probably also the reason, why Abū Maᶜshar discussed comets, when he was asked again for *al-kayd*, see 2nd paragraph cited above. The previously earliest mention of *al-kayd* in an Arabic text was in Ibn Hibintā's *al-Mughnī* written A.D. 829: *al-kayd* "is one of the stars with tail; it appears once every hundred years and travels retrogradely, like the lunar nodes ..."[53] The period given above for the lunar nodes (2.5 degrees per year, i.e. 144 years) is roughly twice the orbital period of comet 1P/Halley.

In the second sentence of the first paragraph, Abū Maᶜshar quotes the opinion of other people, maybe scholars, on what *al-kayd* may be, namely a [*nebulous*] *spot* (*laṭkhatun*). The next possibility of *al-kayd* given (*some people have argued, it would be a bright southern star* and that *scientific astronomers say it is something which produces traces* [*yu'aththiru*]), may point to a (super-)nova, as the Arabic *athar* has been used for super-novae, e.g. SN 1006[54], but in an astrological context, it can also mean *portent* or *trace* or *effect*. Another possibility given (*the nodes of certain spheres*) seems to be the correct one, given the origin of the word *al-kayd*, but it was not favoured by Abū Maᶜshar. In the second paragraph, Shādhān considered that *al-kayd* would be identical to (one of) *the twelve nayāzik*, which was not supported by Abū Maᶜshar.

The discussion in the second paragraph leads us back to *al-kayd*: Shādhān mentions a *bright southern star* in the sign of *Aquarius,* brighter than Mercury. Abū Maᶜshar identifies it with the star called *The Head of Orion,* while Shādhān mentions that the Persians called it *al-bārī* (a corrupt middle-Persian name; later, the Persian *al-bār* was used for α Aur). We would like to point out that the star called *The Head of Orion* in Ptolemy's star catalog is very faint (λ *Orionis* with 4th mag). It is possible that the connection to the discussion about *al-kayd* (maybe a nebulous spot as one possibility given) is due to the fact that Ptolemy describes the star discussed above as *nebulous* (λ Ori forms a small triangle with φ$^1$ und φ$^2$ Ori, which can appear unresolved and then like a nebulous spot, today known as a star forming gas cloud), while Shādhān describes the star as being as white as or larger than Mercury, maybe referring to a whitish extended nebula.

The word *nayzak* (pl. *nayāzik*) can stand for one of the twelve original Greek types of comets, but it is of Persian origin and means *spear* – the comet *looking like a spear*[55], or it can stand for *comet(s)* in general, or – even more general – for *transient celestial object(s)* or *guest star(s)*.[56] Abū Ma cshar also used *al-dhu'āba* for the comet (*lock of hair*) in the text above. We translated the word *kawākib* in the 2nd text as *celestial objects*, as it can mean either *stars* or *planets*.

It is not clear, who the astronomer or astrologer *al-Jaiyānī* is, who is mentioned at the beginning of the second paragraph. The name could point to *Jaen*, a Spanish town, but it is highly doubtful whether an astronomer from Arab Spain (al-Andalus, Arabic from A.D. 711 to 1492) would be known to Abū Ma cshar in Baghdad, Iraq, already in his life-time (he died A.D. 886 at an age of roughly 100 lunar years) or to his student Shādhān. There is no Spanish astronomer or other Spanish person known with this name. If the person would not be contemporary with Abū Ma cshar and/or Shādhān, one would have to consider whether this part of the text was added later by a copyist.

There is, however, a person known as *al-Jaihānī*, who lived in Persia from around A.D. 900 to at least A.D. 978; he was known to be interested mostly in geography, but also in astronomy; as a vizier, he invited foreign scholars to ask them about their countries and astronomy, e.g. the local altitude of Polaris; one of his known friends was Abū Zayd al-Balkhī[57], i.e. from Balkh like Abū Ma cshar – hence a possible connection between *al-Jaihānī* (or al-Jaiyānī ?) and Abū Ma cshar. According to Miguel, the reports about the person named *al-Jaihānī* refer to two different persons, father and son[58], so that the father may have lived already in the ninth century A.D., contemporary with Abū Ma cshar. The two names (*al-Jaihānī* and *al-Jaiyānī*) differ only by one letter; while it may be unlikely for a copyist to mix up these two Arabic letters (*h* instead of *y*), because they look quite different in Arabic, the two names still sound similar, so that either a copyist (mis)wrote while the MS was read to him, or the name was wrong from the very beginning (heard and/or written wrong by Shādhān).

We conclude that Abū Ma cshar's arguments were not understandable from the Latin translations, which may have been translations from an incomplete and altered Greek or Hebrew translation of the original Arabic. Pingree numbered the relevant paragraph (the 2nd paragraph as given in Arabic and English) with no. 106 for MS Saib 199, but did not give the Arabic text nor an English translation[59]; he also noticed that this paragraph was partly present in the two Greek MSS and in Federici Vescovini's edition of the Latin MSS[60].

4  Astronomical interpretation of the Arabic text and Abū Ma cshar's observation

In the last part of the second paragraph as cited above (Sect. 3), Abū Ma cshar explains his observation: Abū Ma cshar had observed (*years ago*) a *western* (evening) *comet*. Venus was seen in close conjunction with the comet, actually Venus was apparently seen as projected onto the comet's tail (*Venus was standing near it in its body*). Venus and the comet apparently did not move relative to each other from night to night (*not moving away from it*). Both Venus and the comet were clearly seen as different objects (*its light did not mix with its light and there was no blending between the two lights*). There is no mention of colour contrary to the English translation[61] of a (wrong) Latin translation of this work.

Abū Maʿshar then also explains that other colleagues have made similar observations of comets with both Jupiter and Saturn (*more than one told me that they saw Saturn in a similar manner with certain comets, and also Jupiter*).

Then, the interpretation of Abū Maʿshar is as follows: "I could conclude that it [the comet] stood above of Venus, because the light of Venus was unimpaired for the eye." Hence, Abū Maʿshar observed Venus in (or actually projected onto) the tail of a comet. With the wording the *light of Venus was unimpaired*, Abū Maʿshar probably meant that Venus was seen with the same brightness while in (projected onto) the cometary tail as before and/or afterwards, i.e. that the (presumable or expected) extinction due to the cometary tail (if Venus would be behind the tail) was negligible. He then concluded that Venus would be in front of the comet – otherwise, the light of Venus would have been partly absorbed. Since other astronomers also observed Jupiter and Saturn in (projected onto) cometary tails, Abū Maʿshar concluded that comets are beyond (above/behind) the planets, i.e. in particular outside the Earth atmosphere and outside the sphere of the Moon.

He says specifically: "comets (*nayāzik*) are not celestial objects, which orbit in normal spheres, but they are celestial objects, which are formed [or: orbit] in an orbit different from the orbits of the seven [planets]. It appears to me most probable that they are formed [or: orbit] in the zodiacal sphere."

It is clear that Abū Maʿshar considered comets to have distances larger than the planets. He places their orbits and distances to the stars in the sphere of the Zodiac, i.e. in the sphere of the fixed stars. While the conclusion drawn by Abū Maʿshar, that comets are not sub-lunar, is correct, his argument is not correct: Cometary tails are optically thin, so that more distant planets and stars can be seen *unimpaired* through them.

Abū Maʿshar may have known that sometimes (also normal fixed) stars are seen in (i.e. behind) cometary tails, but that the brightness of those stars would also not be affected, so that he would need to conclude that those comets are behind those stars. And indeed, he placed comets in the sphere of stars.

Stars (and planets) with very small angular separation near a bright cometary head (whether projected onto the tail or not) may not be visible anymore; this is then not due to extinction, but due to the large brightness difference – a problem of dynamic range, i.e. not only brightness difference and not only separation: The closer *and* fainter a faint object (like a background star) to the bright (foreground) comet (head), the more difficult it is to detect the fainter object. Maybe this is what was observed by Abū Maʿshar: He may have noticed that a faint star was not detectable anymore while in close conjunction with the brightest parts of a comet (head), while Venus remained unimpaired even in close conjunction to a comet (head). Even in this case, his argument would not be correct: From a very close conjunction of Venus (even if remaining *unimpaired*) and a comet, one cannot conclude that the comet is behind Venus. The problem of dynamic range mentioned before applies to both background and foreground objects.

## 5  Planet – comet conjunctions A.D. 750 to 886

We can now consider which comet was observed by Abū Ma$^c$shar in close conjunction with Venus – and which comets were observed by the other astronomers in close conjunction with Jupiter and Saturn. Since the observers may have misidentified the planet(s), we consider all planets visible for the naked eye.

Let us first discuss which time period to consider: Abū Ma$^c$shar himself died A.D. 886 Mar 9 (272 Hijra). He also mentioned that other astronomers had seen Jupiter and Saturn in close conjunction with a comet, but he does not specify the times nor names of those astronomers. They could have been his teachers, e.g. al-Kindī, whose work is mostly lost. The earliest possible time of observation may therefore probably not be before roughly A.D. 750, when Arabs started to translate Greek texts and to study astronomy. Therefore, we consider the period from A.D. 750 to 886.

Only one comet has a known orbit with sufficient precision[62] for many centuries including the time from about A.D. 750 to 886, so that we can compare it with the planets, namely 1P/Halley. We give the orbital parameters used in Table 1.

**Table 1.** Orbital elements of 1P/Halley for the A.D. 836/837 perihelion (Ref. 62).

| Parameter | SAO | Yeomans & Kiang | Landgraf |
|---|---|---|---|
| Epoch | 837 Mar 10.0 | 837 Mar 10.0 | 837 Mar 10.0 |
| Eccentricity  e | 0.96781 | 0.9678055 | 0.96780525 |
| Apsios  q/[au] | 0.58323 | 0.5823182 | 0.58243687 |
| Perihelion  $T_0$ | 837 Feb 28.27 | 837 Feb 28.27 | 837 Feb 28.48165 |
| Long. asc. node  $\Omega$ | 44.930 | 44.21516 | 44.23947 |
| Equinox | J2000.0 | B1950.0 | B1950.0 |
| Argument of per.  $\omega$ | 100.101 | 100.08403 | 100.09354 |
| Inclination  i | 163.447 | 163.44258 | 163.43718 |

We have then used the software *Cartes du Ciel* V3.1 and *The Sky* V6.0 to calculate the ephemeris to compare the orbits of 1P/Halley[63] with the planets for its perihelion passages in A.D. 760 and 836/837.

Indeed, we found that Jupiter was most likely once observable in the tail of 1P/Halley, namely on A.D. 836 December 19, see Fig. 3: 1P/Halley may have occulted Jupiter, which was at -2.4 mag, and the tail of comet 1P/Halley was considered to have a length of 1.25 degree (according to *Cartes du Ciel* assuming average solar and cometary activity). Given the uncertainties in the orbit[64], we cannot be absolutely sure about the position of 1P/Halley relative to Jupiter. It is well possible (but not certain) that it is this conjunction that is mentioned by Abū Ma$^c$shar to have been observed by some other astronomer(s).

However, the Chinese astronomers started to observe 1P/Halley on A.D. 837 Mar 22[65], and also in an indirect report from al-Kindī, there are no earlier observations mentioned[66]. It is still possible that records about such earlier observations are missing. It would have been possible to discover the comet very close to the bright planet Jupiter on A.D. 836 Dec 19 just because of its close separation from Jupiter.

Let us consider whether comet 1P/Halley was bright enough on Dec 19: The apparent brightness m (in magnitudes, or mag) of a comet as observed from Earth can be calculated by

m = M + 5 log (d) + 2.5 n log (r)

with absolute brightness M at 1 au, distance d in au between Earth and comet (well known due to the orbit), activity parameter n depending on solar and cometary activity (n=2 for pure reflection), and distance r in au between comet and Sun. For the absolute brightness, values between 3.71 and 5.50 mag have been suggested for 1P/Halley, and for the activity parameter, values from 2.70 to 5.15 have been proposed.[67] For these ranges of parameters and the known orbit, we can estimate the apparent brightness of 1P/Halley for A.D. 836 Dec 19 to be between 5.6 and 7.9 mag (3σ range being 3.7 to 10.3 mag), i.e. possibly detectable. The Chinese have detected 1P/Halley until at least A.D. 837 Apr 28[68], and indeed, for all six parameter combinations suggested, 1P/Halley would have been brighter than 6th mag at the end of April 837.

Furthermore, there was a close conjunction of 1P/Halley with Mars on A.D. 837 Apr 7, Mars being 3.8 degrees away from the comet; at this time, the comet was indeed observed by both the Chinese[69] as well as al-Kindī[70], but the planet is not mentioned in their reports; according to the best orbit, Mars was not inside the comet tail (Fig. 4), but it could have been in the tail given the orbital uncertainties[71], typically several degrees.

Around A.D. 837 Apr. 7, Saturn may also have been located on the tail of 1P/Halley, even with a separation of 77 degrees (Fig. 4): The Chinese reported for A.D. 837 Apr 7 a tail of "two zhang long and three chi wide", i.e. 20 degrees long and 3 degrees wide, and a few days later (Apr 13/14), the tail had grown to a length of eight zhang (80 deg)[72]. The fact that the tail does not appear that long in Fig. 4 is due to the fact that the software used, *Cartes du Ciel*, assumes average solar and cometary activity. We have seen above, that there is evidence for larger than average values for solar and/or cometary activity given the detection on A.D. 837 Apr 28. Furthermore, it was shown that also solar activity was larger than average in the years of A.D. 836 and 837 (and also around A.D. 760, the previous perihelion), seen by strongly enhanced auroral activity and a drop in radiocarbon[73]. Hence, it is well possible that 1P/Halley was brighter than average and that its tail was longer than average, consistent with the historical reported tail length (80 degrees), so that Saturn may have been located just inside (actually projected behind) the tail.

The fact that the tail was observed to be much longer than expected for average solar and cometary activity may be considered additional evidence for strong solar (or cometary) activity at that time.

We could also find some additional (not that) close conjunction of Venus with 1P/Halley on A.D. 760 Mar 2 and Jun 12 and on A.D. 837 Jan 14. Since Abū Ma'shar mentioned that he had observed such a conjunction (with Venus) himself, he could not have meant those in A.D. 760. For A.D. 837 Jan 14, Venus was as close as 3.5 degrees from 1P/Halley, Venus had -4.1 mag and the tail of the comet had a length of 2.3 degrees (Fig. 5a). With the same arguments and assumptions as above, 1P/Halley may have been visible on Jan 14 at about 5th magnitude. As also seen in Fig. 5a, Venus may not have been in the tail of the comet. However, within the orbit uncertainties[74], we cannot exclude that Venus was seen in the tail at this occasion (the $1\sigma$ positional uncertainty for the comet is shown in Fig. 5a as circle, Venus lies just outside this circle). At this conjunction, comet 1P/Halley was at 1.6265 au distance from Earth and Venus at 1.0406 au, so that indeed, the comet was behind Venus - as Abū Ma$^c$shar may have thought.

We can otherwise exclude close conjunctions with the planets and 1P/Halley for the period studied. The conjunctions reported by Abū Ma$^c$shar may of course be related to other comets, not 1P/Halley.

There is an observation of a conjunction between a planet identified as Venus and a comet reported from Korea in the Chronicle of Silla for A.D. 867/868, i.e. within the life-time of Abū Ma$^c$shar: "During the 12th month of the seventh year of Kyongum Wang [A.D. 867 Dec 30 to 868 Jan 28] a guest star trespassed against Venus" (Sanguk Sagi 11/5).[75] This comet is not mentioned in Pankenier et al.[76], because it does not satisfy their selection criteria, as it is not clear from the report whether it was a comet or nova.

The above date (A.D. 867/868) would then also be quite close to the A.D. 866 Apr 19 passage of *al-kayd* as descending lunar node through the vernal point[77] – possibly the reason why Abū Ma$^c$shar combined the comet-Venus conjunction with *al-kayd* in his memory. However, in that month (A.D. 868 Jan), Venus was visible in the East in the early morning around -4.2 mag, while Abū Ma$^c$shar reported to have seen Venus in close conjunction with a western, i.e. evening comet (*al-dhu'āba al-maghribīya*). (In the quotation by Licetus[78] as given in Sect. 2, the year A.D. 844 was given for the observation by Abū Ma$^c$shar. In the text of Abū Ma$^c$shar himself, however, there is no indication for this year (*years ago*). While it is possible that the text, on which the Latin translation consulted by Licetus is based, is different from Ankara Saib 199, it was noted before that there was a strong confusion about the year of this (or some other nova or comet) observation.[79]) It is, however, possible that both the Koreans and Abū Ma$^c$shar misidentified the planet as Venus, but actually observed a different planet close to a comet.

In addition to 1P/Halley, there are three comets with known orbital solutions only for one particular perihelion passage between A.D. 750 and 886, namely those C/770 K1, C/817 C1, and C/868 B1.[80] We have compared their orbits with the positions of the planets, too. According to the orbital solution of the comet of A.D. 868[81], planet Mars with about 0.9 mag apparent brightness may have occulted the comet (brightness 2.7 mag for typical activity, n=4) on A.D. 868 Jan 27 (Fig. 5b), located in the evening in the west, 41 degrees altitude during the closest approach. It may still be true that Venus was close to a comet between A.D. 867 Dec 30 to 868 Jan 28, as reported from Korea, even though the orbital solution[82] does not support this report, because it is well possible that the reports about the comet of A.D. 867/868 are actually about two different comets[83]. Alternatively, it is also possible that the Koreans have thought or reported that it was Venus, but in fact it was Mars. Maybe, the same error was committed by Abū Ma$^c$shar (reporting a western, i.e. evening comet).

Interestingly, the planets Venus and Mercury were close to the previous comet of A.D. 868 on A.D. 868 Mar 27, both separated from the comet by only 1 degree, but this was at an altitude of only 2.5 degrees, the comet may have been invisible at m=6.8 mag for typical activity, and the conjunction was in the East (Fig, 5c). It could be, though, that the Korean report cited above, which connects the comet of A.D. 868 with Venus, is just misdated (to A.D. 868 Jan), but really refers to the close conjunction on Mar 27. At this conjunction, the comet was at 1.4424 au distance from Earth and Venus at 1.4410 au, so that indeed, the comet was slightly behind Venus - as Abū Ma$^c$shar may have thought.

Finally, the comet of A.D. 770 had a close conjunction with Mars on A.D. 770 Apr 28 with a separation of 7 degrees (Fig. 5d). If it was mistaken for Jupiter or Saturn, it could have been one of the observations mentioned by Abū Ma$^c$shar. Comet brightness (4.5 mag) and tail length are plotted in Fig. 5d for average cometary and solar activity, but the Sun was more active than on average in around A.D. 770.[84]

Furthermore, for some 30 additional comets reported in historical documents from A.D. 750 and 886, it is specified in which constellation(s) they were seen[85], so that we could compare all those reports with the positions of the planets at those dates to find possible close conjunctions. We then found three more possible cases: (i) Chinese observations report a comet in Orion for A.D. 852 Mar and Apr in the western evening sky with a long tail of 50 degrees[86], which should be pointing away from the Sun, i.e. upwards on the sky towards the east; and indeed, Jupiter at -1 mag was located nearby towards that direction between Virgo and Cancer. (ii) Another Chinese comet is reported for A.D. 857 Sep and Oct in Scorpius[87] in the southwest; with the Sun in the west, the tail should point to the east, and indeed Jupiter was located towards the east nearby between Ophiuchus and Sagitarius at -1 mag. (iii) For A.D. 864 Apr and May, a comet is reported (from China and Japan) for Aries[88] in the west; its tail was 3 degrees long and should point to the east, and indeed, Mars and Venus were located nearby towards the east in Taurus. It may well be possible that Abū Ma$^c$shar was reporting about any of those (close ?) conjunctions, but we cannot confirm it without further details.

Another close conjunction between a planet (Venus) and a comet is reported for A.D. 907 Apr 9 observed in Japan, for a comet with a 30 degree long tail that "trespassed against Venus"[89], the planet was then indeed in the western evening sky, but this was after the death of Abū Ma$^c$shar.

6  Summary

From the fact that Venus was seen in the tail of a comet, Abū Ma$^c$shar concluded that the comet was located behind Venus – obviously (incorrectly) considering the comet tail to be optically thick. While some part of the conclusion is correct (that comets are outside the Earth atmosphere and behind the moon), the argument itself is not justified, because cometary tails are optically thin. Due to similar observations with Jupiter and Saturn, he concluded that comets are located in the sphere of the stars. This was seen as a clear contradiction with Aristotle's *Meteorology*. Abū Ma$^c$shar said *I am not happy with any of those* (opinions of other people), while in the English translation of a Latin translation we find: "The philosophers say, and Aristotle himself, that comets are in the sky in the sphere of fire ... But they all have erred in this opinion."

The conclusion in *Abū Ma$^c$shar in Shādhān* (late ninth century A.D.), to place comets beyond Venus, clearly ante-dates the same conclusion by Tycho Brahe (for the comet of A.D. 1577) some seven centuries later – and Tycho Brahe quotes Abū Ma$^c$shar for his much earlier finding.

While we cannot be sure which planet–comet conjunctions were observed and mentioned by Abū Ma ͨshar, we found a few good candidates, namely conjunctions of 1P/Halley with Jupiter (Halley possibly visible) and Saturn (near the end of a very long tail) in A.D. 836/837 as well as a conjunction of comet C/868 B1 with Mars, where planet Mars was misidentified as Venus by the Koreans and maybe also by Abū Ma ͨshar.

Even though the argument made by Abū Ma ͨshar is not justified, it is nevertheless highly important to note that he considered this possibility – contrary to the common interpretation of Aristotle. This is but one more example where Arabic scholars did not only use and translate Greek texts, but where they also questioned them. It may well have helped Tycho Brahe much later to come to a similar result.

Like Tycho Brahe, Abū Ma ͨshar may have seen the problem that comets appear to cross the planetary spheres. Contrary to Tycho Brahe, Abū Ma ͨshar solves this problem by placing the comets behind the planets, among the stars. This solution may have been motivated by the observation of Saturn (the most distant planet know at that time) in the tail of a comet, so that it was concluded (incorrectly) that the comet was behind Saturn, and/or by another observation that also the light of star(s) remained unimpaired when in close conjunction with a comet. However, comets behind Jupiter and Saturn cannot be detected by the naked eye.

These further conclusions were not known to Tycho Brahe, who just quotes Adam Ursinus following Cardanus: "Albumasar said: A comet was seen above Venus; it is therefore not in the sphere of the [four] elements." This was consistent with the findings of Tycho, who concluded that comets are not sub-lunar. If he would have known that *Abū Ma ͨshar* had placed comets outside the solar system, as specified in the Arabic text, he might not have quoted him, but that part of the quotation was lost during the transmission through Latin.

After the relevant quotation (*Albumasar said: A comet was seen above Venus*), Tycho Brahe then continues as follows (translated by us):

"This statement, I think, was brought up suitably in that citation, and he [Ursinus] correctly reminds us that some basics of the oldest astronomers on these matters should be reconsidered and be compared more carefully with this phenomenon [new star of 1572]. If he [Ursinus] would have done so, he would not have listed this star [of 1572] as comet. Nevertheless, if he [Ursinus] thinks, that this opinion of Albumasar is superior to that of Aristotle, and if he thinks that all comets belong to the sky [supra-lunar], and if he has therefore considered this new star [of 1572] to be in that heavenly zone [outside the solar system] (which he did not do publicly), then he deserves patience for unsuitably listing this [supernova 1572] as comet."

Likewise, we can state: If Abū Ma ͨshar thinks that all comets belong to the sky [supra-lunar], and if this has helped Tycho Brahe some 700 years later to come to a similar conclusion, then he deserves patience for unsuitably considering a comet tail as intransparent.

Acknowledgements.
We would like to dedicate this paper to Prof. Dr. Fuat Sezgin, who found the Ankara Saib 199 manuscript and the two relevant paragraphs; we use a copy of the MS from his library here. We thank the staff at the Institut für Geschichte der Arabisch-Islamischen Wissenschaften, Frankfurt am Main, Germany, the Saib Library at U Ankara, Turkey, and the Süleymaniye Library in Istanbul, Turkey, for their assistance in searching for MSS. The Bibliotheque nationale de France in Paris, France, and the Millī Library in Teheran, Iran, also provided digital copies of (other) MSS; we thank Pouyan Rezvani (Teheran, Iran) for providing MS Millī and for searching for the MS in the Khunjī Library, both Teheran, Iran. We would also like to thank B. Dinçel (U Jena), M. M. Serim, S. Yerli (METU Ankara), H. Serim (Ankara), A. Farhan (Boĝaziçi U Istanbul), T. Ak (U Istanbul), and T. Seidensticker (U Jena) for their help with the MSS. We consulted several of the literature listed in the library of the Institut für Geschichte der Arabisch-Islamischen Wissenschaften, Frankfurt am Main, Germany, and would like to thank F. Sezgin also for pointing us to *al-Jaihānī*. RN would also like to thank D.L. Neuhäuser for many good suggestions.

---

[1] E.g. M. Weichenhan, *Ergo perit coelum, Die Supernova des Jahres 1572 und die Überwindung der aristotelischen Kosmologie* (2004, München) Franz Steiner.

[2] Tycho Brahe, *Astronomiae Instauratae Progymnasmata. Quorum haec Prima Pars. De restitutione motuum Solis et Lunae Stellarumque inerrantium Tractat et praeterea de admiranda Nova Stella anno 1572* (1602, Uraniburgi Daniae, Pragae Bohemiae), p. 783 citing A. Ursinus, *Prognosticatio Auff das Jhar nach der Geburt Jesu Christi unseres Heilands, M.D. LXXIIII. Beyneben einer kurtzen Beschreibunge des erschienenen Cometens im 1572. und 1573. Jhare*, p. 781-783 citing H. Cardanus, *Aphorismorum Astronomicorum Segmenta* VII, Opusculum incomparabile, appended to the edition of 1547, H. Cardani Libelli 5, p. 217; see also W. Hartner, Tycho Brahe et Albumasar, *La Science au seizième siècle* (1960, Paris), p. 137-150.

[3] F. Sezgin, *Geschichte des Arabischen Schrifttums*, VII: Astrologie, Meteorologie und Verwandtes (1979, Leiden), p. 139-151.

[4] D. Pingree, Historical Horoscopes, *Journal of the American Oriental Society* 82, p. 487 (1962); D. Pingree, Abū Maᶜshar al-Balkhī, Jaᶜfar ibn Muḥammad, *Dictionary of Scientific Biography* I (1970, New York), p. 32-39.

[5] Sezgin, op. cit. (ref. 3), p. 139.

[6] Sezgin, op. cit. (ref. 3), p. 139; D.M. Dunlop, The Mudhākarāt fī ᶜIlm an-Nujūm, *Iran and Islam*, C.E. Bosworth, Ed., (Edinburgh, 1971), p. 230 citing the *Fihrist* by Ibn al-Nadīm, and Ibn Khallikān, biographer of Abū Maᶜshar.

[7] Sezgin, op. cit. (ref. 3), p. 141-142.

[8] F. Sezgin, *Geschichte des Arabischen Schrifttums*, VI: Astronomie (1978, Leiden), p. 156-157.

[9] Sezgin, op. cit. (ref. 3), p. 203-216.

[10] B.R. Goldstein, Evidence for a supernova of A.D. 1006, *Astronomical Journal* 70, 105-114 (1965).

[11] Goldstein, op. cit. (ref. 10).

[12] Licetus, *De novis astris et cometis* (1623, Venice).

[13] Goldstein, op. cit. (ref. 10), p. 110.

[14] L. Thorndike, Albumasar in Sadan, *Isis* 45 (1954), p. 22-32.

[15] Thorndike, op. cit. (ref. 14), p. 29.

[16] G. Federici Vescovini, La versio Latina degli Excerpta de secretis Albumasar di Sadan, Una edizione, Archives d'Histoire Doctrinale et Litteraire du Moyen Age (1998), Vol. 65, p. 273-330.

[17] Thorndike, op. cit. (ref. 14), p. 29.

[18] Thorndike, op. cit. (ref. 14).